\documentclass{article}

\usepackage[final,nonatbib]{neurips_2024}

\usepackage[utf8]{inputenc} 
\usepackage[T1]{fontenc}    
\usepackage{hyperref}       
\usepackage{url}            
\usepackage{booktabs}       
\usepackage{amsfonts}       
\usepackage{nicefrac}       
\usepackage{microtype}      
\usepackage[svgnames,dvipsnames]{xcolor}         

\usepackage{graphicx}
\usepackage{doi}
\usepackage{multirow}
\usepackage{tikz}
\usetikzlibrary{decorations,arrows,shapes}
\usetikzlibrary{backgrounds}
\usepackage{amsmath}
\usepackage{caption}
\usepackage{subcaption}
\usepackage{pifont}
\newcommand{\cmark}{\ding{51}}
\newcommand{\xmark}{\ding{55}}
\usepackage{algorithm}
\usepackage{algpseudocode}
\algrenewcommand{\algorithmiccomment}[1]{\hfill \textcolor{ForestGreen}{\# #1}}
\algnewcommand{\LeftComment}[1]{\textcolor{ForestGreen}{\# #1}}
\usepackage{ragged2e}

\title{QuickBind: A Light-Weight And Interpretable Molecular Docking Model}

\author{%
  Wojtek Treyde\thanks{Present address: Department of Chemistry, University of Oxford, Oxford, UK.} \\
  Department of Systems Biology\\
  Columbia University\\
  New York, NY \\
  \texttt{wojtek.treyde@sjc.ox.ac.uk}
  \And
  Seohyun Chris Kim\\
  Department of Systems Biology\\
  Columbia University\\
  New York, NY
  \AND
  Nazim Bouatta \\
  Department of Systems Biology\\
  Harvard Medical School\\
  Boston, MA
  \And
  Mohammed AlQuraishi\\
  Department of Systems Biology\\
  Columbia University\\
  New York, NY \\
  \texttt{m.alquraishi@columbia.edu} \\
}

\begin{document}
\maketitle

\begin{abstract}
Predicting a ligand's bound pose to a target protein is a key component of early-stage computational drug discovery. Recent developments in machine learning methods have focused on improving pose quality at the cost of model runtime. For high-throughput virtual screening applications, this exposes a capability gap that can be filled by moderately accurate but fast pose prediction. To this end, we developed \textsc{QuickBind}, a light-weight pose prediction algorithm. We assess \textsc{QuickBind} on widely used benchmarks and find that it provides an attractive trade-off between model accuracy and runtime. To facilitate virtual screening applications, we augment \textsc{QuickBind} with a binding affinity module and demonstrate its capabilities for multiple clinically-relevant drug targets. Finally, we investigate the mechanistic basis by which \textsc{QuickBind} makes predictions and find that it has learned key physicochemical properties of molecular docking, providing new insights into how machine learning models generate protein-ligand poses. By virtue of its simplicity, \textsc{QuickBind} can serve as both an effective virtual screening tool and a minimal test bed for exploring new model architectures and innovations. Model code and weights are available at this \href{https://github.com/aqlaboratory/QuickBind}{GitHub repository}.
\end{abstract}

\section{Introduction}
Small organic molecules ("ligands") are a major class of drugs that act by binding protein targets, thereby affecting their functionality and interfering with the molecular pathways of diseases. Their distinct advantages, including ease of synthesis, administration, and cell permeability, render them indispensable as a pharmaceutical modality. In early-stage drug discovery, structure determination of protein-ligand complexes is a critical scientific tool, as it can provide an understanding of the molecular determinants of binding and enable further optimization of drug affinity and selectivity. It is however bottlenecked by costly and cumbersome experimental procedures, which computational "molecular docking" promises to overcome. When supplemented by an estimate of the strength of the binding interaction, computational tools can be used to virtually screen very large spaces of drug-like molecules \cite{ChemicalSpace} for viable drug candidates that can serve as starting hypotheses for subsequent experimental investigation and development \cite{Ferreira.2015, Lyu.2019}. Existing computational methods achieve high-quality predictions at the cost of increasingly long runtimes, caused, in the case of conventional physical methods, by the need to sample numerous binding locations and poses, and in the case of machine learning (ML)-based methods by the complexity of the underlying neural computations that implicitly do the same.

Modern methods can be largely divided into molecular docking and co-folding. In molecular docking, an approximate protein structure is assumed to be known, whereas in co-folding both protein and ligand structures are predicted from scratch. Although a new development, co-folding has become the focus of much recent research activity, including \textsc{RoseTTAFold All-Atom} (\textsc{RFAA}) \cite{RFAA}, \textsc{NeuralPLexer} \cite{NeuralPLexer}, \textsc{Umol} \cite{UMol}, and \textsc{AlphaFold 3} (\textsc{AF3}) \cite{AF3}. Nonetheless, for drug discovery campaigns against a known and potentially well-studied protein target, it is often unnecessary to predict protein structures independently for every ligand, making molecular docking an attractive alternative given its higher speed (due to the assumed rigidity of the protein).

ML-based docking methods can be further divided into targeted docking, which requires specifying the approximate binding pocket, and blind docking, which does not. The first ML method to tackle the latter is \textsc{EquiBind} \cite{EquiBind}, which predicts the isolated, bound conformation of the ligand and uses a keypoint alignment mechanism to identify the rotation and translation needed to dock the ligand into the binding pocket. Subsequent methods employ more complex architectures. In particular, \textsc{TANKBind} \cite{TANKBind} and \textsc{E3Bind} \cite{E3Bind} first partition the protein into functional blocks using P2Rank \cite{Krivak.2018}, predict the interaction of the given ligand with each block, and choose the final pose based on the predicted binding affinity (\textsc{TANKBind}) or confidence score (\textsc{E3Bind}). Both architectures include components inspired by \textsc{AlphaFold 2}’s (\textsc{AF2}'s) Evoformer module \cite{Jumper.2021}. Importantly, \textsc{TANKBind} predicts an intermolecular distance map that is converted into final coordinates by numerical post-optimization, whereas \textsc{E3Bind} operates on the ligand coordinates directly. \textsc{FABind} \cite{FABind} builds upon \textsc{E3Bind} by integrating the prediction of the location of the binding pocket into the main model, such that the whole process becomes end-to-end differentiable. Another leap forward was made by \textsc{DiffDock} \cite{DiffDock}, a diffusion-based generative model that, starting from an input conformer, predicts changes in torsion angles as well as the transformation needed to dock the ligand into the protein. The inductive bias to focus only on relevant degrees of freedom coupled with a generative formulation led to large improvements in accuracy (and increased runtimes). More recently, advances have come from integration of protein language models \cite{Lin.2022}, pretraining techniques geared towards molecular docking tasks, substantial increases in model size, and a shift towards generative approaches \cite{DiffDock, NeuralPLexer, AF3}. In combination, these trends have led to considerable increases in the computational cost of ML-based molecular docking, at levels prohibitive for virtual screening.

\begin{figure}[htpb]
    \centering
    \includegraphics[width=\linewidth]{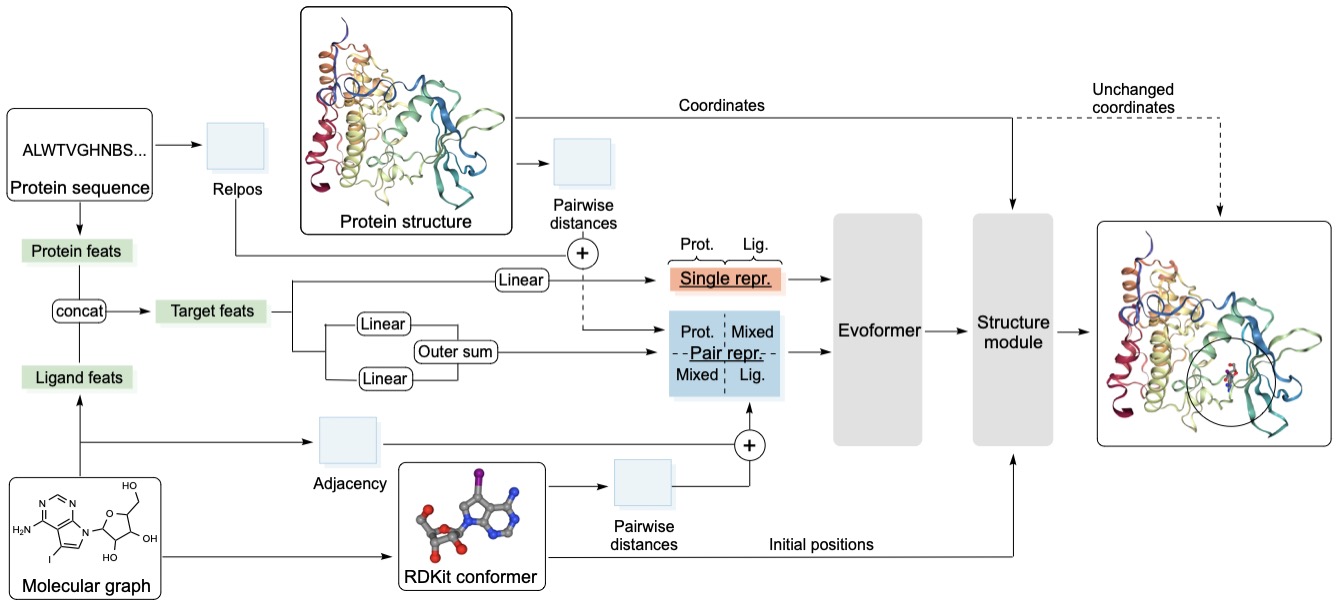}
    \caption{\textsc{QuickBind} architecture. A "single" representation is first constructed by concatenating embedded protein and ligand input features. A "pair" representation is then constructed from linear embeddings of the single representation, pairwise distances (of protein residues and ligands atoms, independently), relative positional encodings of protein residues, and the adjacency matrix of ligand atoms. The pair representation contains a protein and a ligand block, as well as mixed off-diagonal elements. The single and the pair representations are passed through a modified Evoformer stack, before the Structure module uses the updated single and pair representations as well as initial coordinates from an RDKit conformer \cite{rdkit} and protein coordinates from the input protein structure to dock the ligand into the binding pocket.}
    \label{fig:architecture-overview}
\end{figure}

In this work we develop \textsc{QuickBind}, a light-weight method for rigid, blind molecular docking aimed at virtual screening applications by trading accuracy for speed. \textsc{QuickBind} performs well on the PDBBind test set, in particular when only unseen proteins are considered, and is substantially faster than \textsc{DiffDock}. Leveraging the AF2 architecture and problem formulation (\autoref{fig:architecture-overview}), \textsc{QuickBind} reasons over proteins using a residue-level representation in lieu of an atomistic one, permitting fast but implicit accounting of side chain flexibility. To accommodate additional degrees of freedom introduced by small molecule ligands, \textsc{QuickBind} incorporates a new framing strategy for the Invariant Point Attention (IPA) module of AF2. We also include a binding affinity prediction module to facilitate virtual screen applications, a capability typically absent from docking and co-folding methods. We showcase the utility and versatility of \textsc{QuickBind} by predicting binding affinities and structures for multiple important drug targets across various protein families, and investigate the interpretability of our model to better understand the biophysical basis of its predictions.

\section{Methods}
\subsection{Dataset and evaluation metrics}
We train and test \textsc{QuickBind} using the PDBBind \cite{Liu.2017} dataset and additionally assess it using the PoseBusters (PB) Benchmark \cite{Buttenschoen.2023}. PDBBind has been widely used to assess molecular docking methods using a temporal split proposed by \textsc{EquiBind} \cite{EquiBind}. It contains crystal structures and binding affinities for $\sim$20,000 protein-ligand complexes; the training and validation sets comprise 16,379 and 968 complexes, respectively, both published before 2019, while the test set contains 363 complexes published in 2019 or later. There is no ligand overlap between the three partitions. PB contains 428 diverse complexes of unique proteins and drug-like ligands released since 2021, and is therefore disjoint from complexes in the PDBBind training set.

We quantify success based on the percentage of predictions with a symmetry-corrected ligand heavy atom root-mean-squared deviation (RMSD) of less than 2\AA\ ("success rate";  \cite{Alhossary.2015, Hassan.2017, McNutt.2021, Meli.2020}). This criterion is widely used, as the residual deviation from the true bound conformation should not materially impact downstream analyses and optimizations. Evaluating models purely on success rate does not account for chemical and physical validity, but we assess these criteria using the PB suite \cite{Buttenschoen.2023}. Predictions that pass all PB tests and whose RMSD is below 2\AA\ are deemed PB-valid.

\subsection{Model architecture} \label{sec:architecture}
\textsc{QuickBind} adapts the \textsc{AlphaFold 2} \cite{Jumper.2021} architecture to the task of protein-ligand pose prediction (\autoref{fig:architecture-overview}). It takes as input protein sequence and structure as well as the chemical graph of the ligand and its 3D conformer (generated by the RDKit \cite{rdkit}). Inputs are combined to yield a unified first-order ("single") representation and a second-order pairwise ("pair") representation. These are then passed to a modified Evoformer stack that omits column-wise self-attention as multiple sequence alignments are not used. After processing by the Evoformer, a  Structure module takes the updated single and pair representations as input as well as residue and ligand reference frames (using a new framing strategy for ligand atoms described below) and iteratively updates the ligand heavy atom coordinates. The Structure module is modified from \textsc{AF2} so that the IPA module is gated \cite{gatedIPA}, cross attention is performed between the single representations of the ligand and protein during the coordinate update step (Algorithm~\ref{alg:backboneupdate}), and protein atoms are held fixed. \textsc{QuickBind} uses 12 Evoformer and 8 Structure module blocks. Full algorithm details are given in section~\ref{sec:algos}.

\textsc{AF2} uses reference frames to represent the geometry of protein residues: a translation vector corresponds to the coordinates of the C$_\alpha$ atom while a rotation matrix, anchored at these coordinates, is canonically constructed from the N, C$_\alpha$, and C coordinates to encode the orientation of the residue backbone. This representation is natural for linear or branched polymers but for atoms of arbitrary small molecules, there are no canonical frame construction approaches. With the emergence of co-folding methods, two recent models have proposed framing strategies for small molecules \cite{RFAA, NeuralPLexer}, and \textsc{QuickBind} employs its own new approach. First, atom indices are reordered based on the canonical atom ranking of the RDKit \cite{rdkit}. For each heavy atom, its coordinates are treated as the C$_\alpha$ atom and the coordinates of its two adjacent atoms with the lowest indices are treated as the N and C atoms. If an atom has only one bond, we use a dummy atom (Algorithm~\ref{alg:dummyatom}). We then construct atom references frames using the same procedure employed for residue frames.

Using the above framing strategy, we found that during the IPA component of the Structure module, updating the rotation matrices of ligand atom frames improves performance, unlike the approaches of \textsc{NeuralPLexer} and \textsc{RFAA} which only update the translation component and passively reconstruct the frames. We note that while \textsc{AF3} does not use reference frames to reason over protein-ligand complexes, it does use a framing strategy to calculate the predicted aligned error. \textsc{AF3}'s reference frames are constructed using the two closest atoms to a given center atom, and when an atom does not have two neighbors, the frame is ignored. We experimented with an analogous strategy but found that our approach works better.

For the loss function we adopt a modified version of the frame-aligned point error (FAPE) used by \textsc{AF2}. FAPE is computed by performing a set of alignments such that the predicted reference frame of each residue is aligned with its original reference frame in the target structure. FAPE is then the average, clamped RMSD between the predicted and target structures over all alignments. For molecular docking, FAPE can be reformulated as a combination of two components: the RMSD between predicted and target ligand atom positions based on ligand frame alignments and residue frame alignments. The final \textsc{QuickBind} model was trained using this combined FAPE loss, intermediate FAPE losses acting on the outputs of every Structure module block, and a Kabsch RMSD loss corresponding to the ligand RMSD after superimposition of the predicted and target ligand using the Kabsch algorithm \cite{Kabsch}.

Further model details are given in sections~\ref{sec:feats}-\ref{sec:train-details}. Before training the final \textsc{QuickBind} model, aspects of the architecture were optimized using two smaller variants, \textsc{QuickBind-S} and \textsc{QuickBind-M} (section~\ref{sec:hyperparamsscreen}). The results of this hyperparamater screen are summarized in \autoref{tab:hyperparameter-screening}.

\section{Results}
\subsection{Model performance}

\begin{figure}
    \centering
    \includegraphics[width=\linewidth]{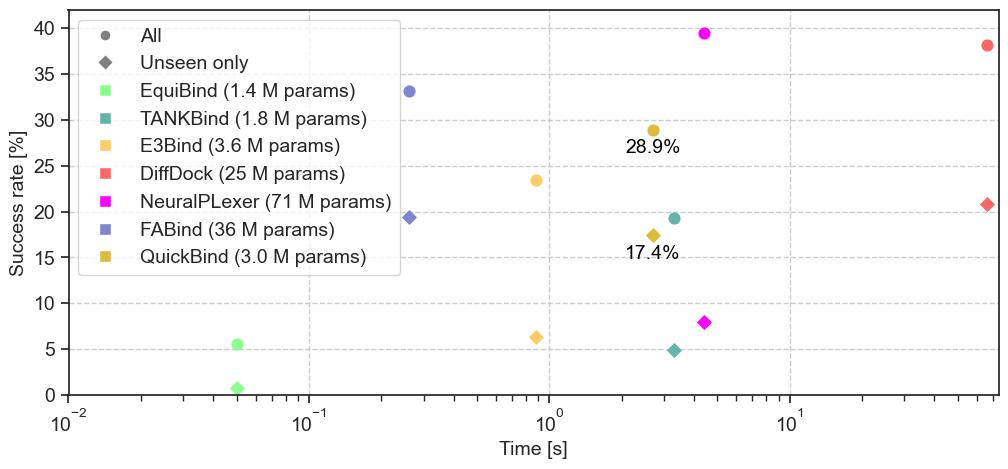}
    \caption{Success rates vs. average runtimes (summed over all complexes) for ML-based rigid docking methods on the PDBBind test set. Success rates are reported separately for all 363 complexes (circles) and for 144 complexes whose proteins are absent from the training and validation sets (diamonds). \textsc{TANKBind} was only evaluated on a subset of 142 unseen proteins by its original authors. Success rates are taken from original publications, except for \textsc{NeuralPLexer}'s success rate on unseen proteins as it was not originally reported. Runtimes do not include preprocessing and were determined on NVIDIA A40 GPUs using scripts provided in each method's respective repository, without batching. \textsc{NeuralPLexer} runtime excludes acquisition of auxiliary inputs (\textit{e.g.,} \textsc{AF2} predictions) while \textsc{TANKBind} and \textsc{E3Bind} runtimes do not include P2Rank segmentation. \textsc{E3Bind}'s runtime was taken from its original publication since authors did not release the model weights and inference code.}
    \label{fig:pdbbind-eval}
\end{figure}

\begin{figure}
    \centering
    \includegraphics[width=\linewidth]{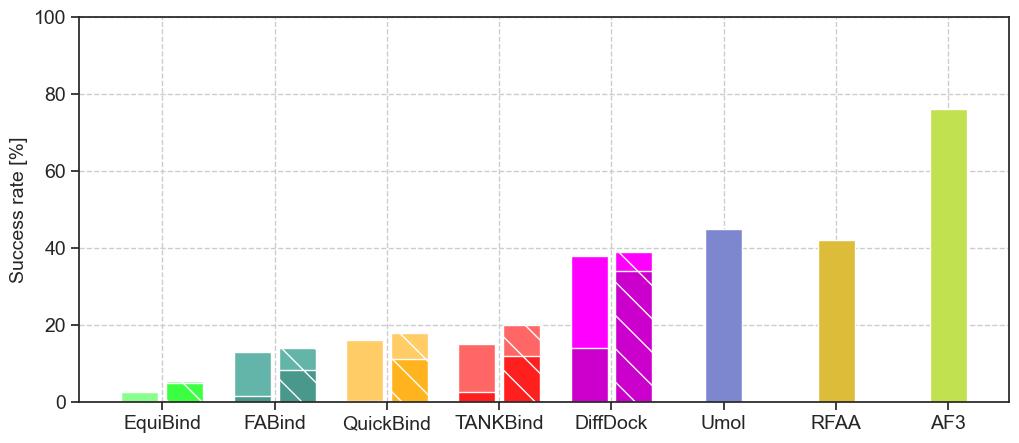}
    \caption{Success rates of ML-based rigid docking and co-folding models on the PB Benchmark, sorted in ascending order by model runtime (left-to-right). Target 7M31 was omitted for \textsc{QuickBind} because it is longer than 2,000 residues. Lighter colors correspond to all predictions while darker colors correspond to PB-valid predictions (docking methods only) and hatched bars to results after energy minimization. Success rates are taken from original publications \cite{RFAA, UMol, AF3, Buttenschoen.2023}, except for \textsc{FABind}'s success rate as it was not originally reported. We found the \textsc{EquiBind} success rates to be 0.9\% (all predictions), 0.2\% (PB-valid predictions), 4.2\% (all predictions after energy minimization), and 3.5\% (PB-valid predictions after energy minimization), in rough agreement with the originally reported values (2.6\%, 0.0\%, 5.5\%, and 4.8\%, respectively).}
    \label{fig:pb-eval}
\end{figure}

We first assess \textsc{QuickBind} on the PDBBind test set (\autoref{fig:pdbbind-eval}) and compare it to other ML-based rigid docking methods including \textsc{EquiBind}, \textsc{TANKBind}, \textsc{E3Bind}, \textsc{FABind}, \textsc{DiffDock}, and \textsc{NeuralPLexer} (employed for blind molecular docking rather than co-folding). We exclude co-folding methods from this comparison as they have not been evaluated on PDBBind. For virtual screening---which can involve billions of molecules \cite{lyu2019ultra}---docking runtimes are a critical consideration. With an average runtime of 2.7s, \textsc{QuickBind} is orders of magnitude faster than traditional docking methods \cite{EquiBind}, \textsc{DiffDock}, and recent co-folding methods \cite{RFAA, UMol, AF3}. Accuracy-wise, \textsc{QuickBind} is middle-of-the-pack, outperforming most ML-based rigid docking methods except for \textsc{FABind}, \textsc{DiffDock}, and \textsc{NeuralPLexer}, particularly when generalizing to proteins excluded from the training set. Excepting \textsc{FABind}, these methods are all considerably slower, and \textsc{FABind} still uses an order of magnitude more parameters. 

Next, we assess \textsc{QuickBind} on the more challenging PB Benchmark (\autoref{fig:pb-eval}) and include in our comparisons recent co-folding methods. \textsc{QuickBind} generally outperforms rigid docking methods (except \textsc{DiffDock} but including \textsc{FABind}) but trails co-folding methods. However, the speed gap between \textsc{QuickBind} and co-folding methods (which take minutes for a single prediction) is even more considerable, making \textsc{QuickBind} a compelling compromise between speed and accuracy. \autoref{fig:examples} shows examples of highly accurate \textsc{QuickBind} predictions.

\textsc{QuickBind} forgoes a post-processing step common to most ML-based methods that enhance the chemical validity of predictions. As a result, \textsc{QuickBind}'s raw predictions do not generally pass the PB chemical and physical plausibility tests, largely due to incorrect bond lengths and angles (\autoref{fig:fail-modes}, see also \autoref{fig:example-a}). As was noted in \cite{Buttenschoen.2023} however, such failures can typically be addressed by running a force field-based energy minimization post-prediction. Doing so does substantially improve the physical validity of \textsc{QuickBind}'s predictions and even slightly improves its success rate (\autoref{fig:pb-eval}).

Given its speed, a natural use case for \textsc{QuickBind} is binding affinity prediction for virtual screening applications. To test this potential, we trained a simple neural network to use \textsc{QuickBind}'s single representations (Algorithm~\ref{alg:binding-affinity-prediction}) to predict protein-ligand affinities. We used the same PDBBind split as for training \textsc{QuickBind} and PDBBind's own affinity data. The resulting model is competitive with other affinity predictors (\autoref{tab:binding-affinity}) despite not having been optimized architecturally or through hyperparameter tuning. This indicates that \textsc{QuickBind} has immediate utility for virtual screening and may be further improved using more advanced top models for affinity prediction.

\subsection{Model interpretability}

Having trained \textsc{QuickBind}, we set out to investigate whether it had learned ligand characteristics relevant for molecular docking. Beyond binding affinity, molecular features such as lipophilicity and cell permeability can influence the outcome of drug discovery programs. These features are strongly linked to the physicochemical properties of ligands, such as hydrophobic surface area and molecular weight, to the extent that expert rules for drug design often explicitly restrict drug-like molecules based on these properties. A model that implicitly encodes them may thus hold promise for drug design beyond accurate pose prediction.

\begin{figure}[htpb]
     \centering
     \begin{subfigure}[b]{0.69\linewidth}
         \centering
         \includegraphics[width=\linewidth]{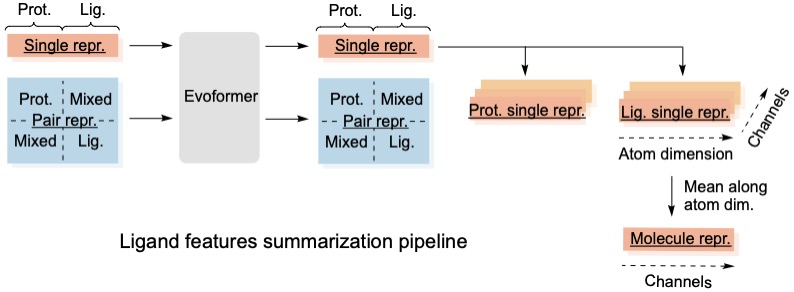}
     \end{subfigure}
     \begin{subfigure}[b]{0.3\linewidth}
         \centering
         \includegraphics[width=\linewidth]{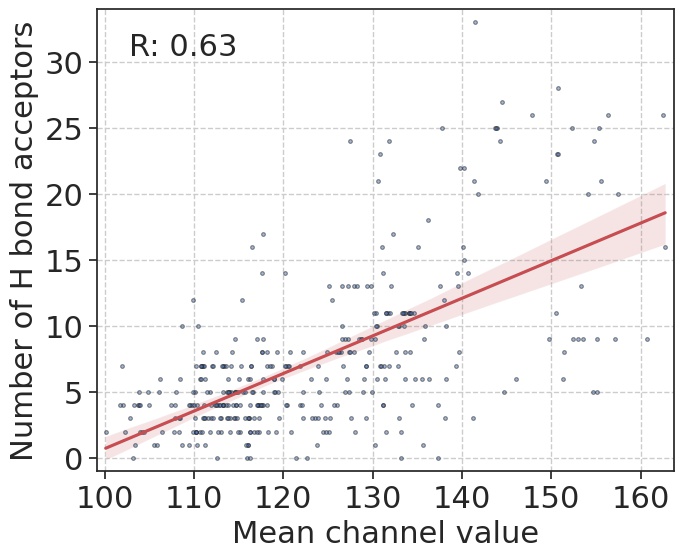}
     \end{subfigure}
     \hfill
     \begin{subfigure}[b]{0.32\linewidth}
         \centering
         \includegraphics[width=\linewidth]{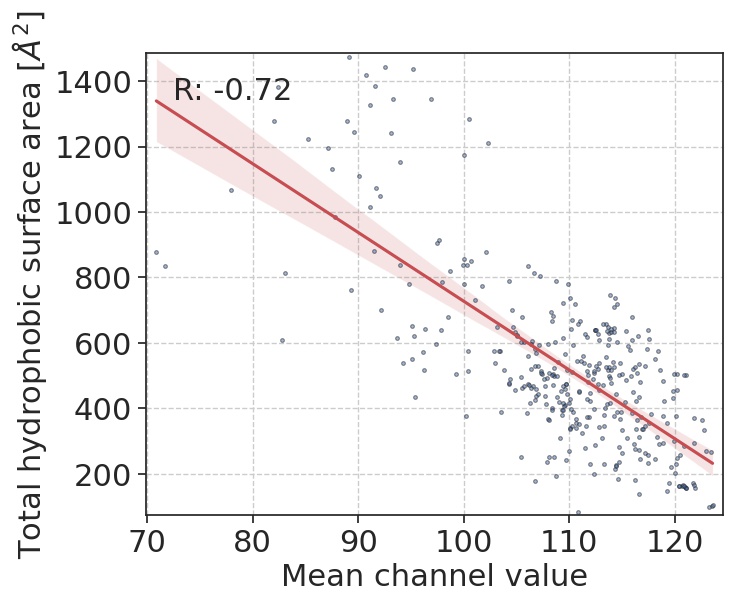}
     \end{subfigure}
     \hfill
     \begin{subfigure}[b]{0.3\linewidth}
         \centering
         \includegraphics[width=\linewidth]{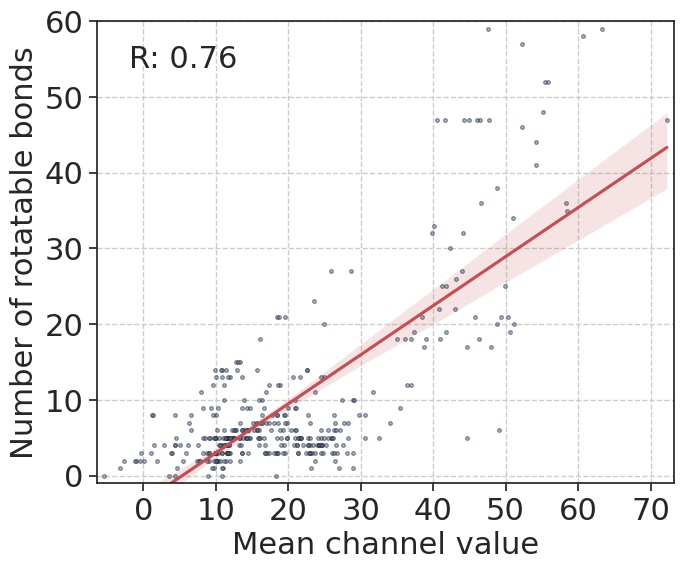}
     \end{subfigure}
     \hfill
     \begin{subfigure}[b]{0.3\linewidth}
         \centering
         \includegraphics[width=\linewidth]{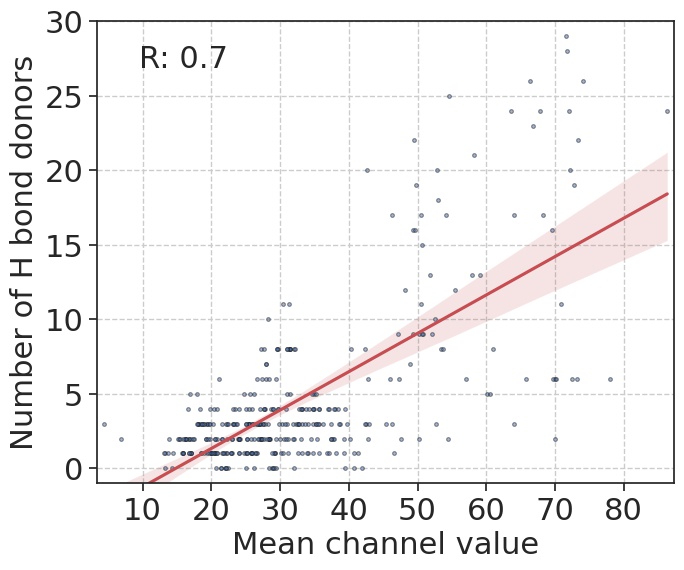}
     \end{subfigure}
        \caption{Interpretable physicochemical properties in \textsc{QuickBind}'s ligand representation. Processing the Evoformer's single representation into separate protein and ligand representations followed by averaging the ligand's atom dimension yields interpretable descriptors that correlate with channel values, including number of H-bond acceptors and donors, total hydrophobic surface area, and number of rotatable bonds.}
        \label{fig:single-rep-interpret}
\end{figure}

To this end, we extracted the single ligand representation from the Evoformer and transformed it into a molecule-level representation by averaging across atoms (\autoref{fig:single-rep-interpret}; pipeline schematic). For every resulting channel value, we computed Pearson's R value for total hydrophobic surface area, molecular weight, number of hydrogen bond acceptors and donors, polar surface area, number of rotatable bonds, octanol-water partition coefficient, and number of aromatic rings; calculations were performed using the RDKit \cite{rdkit} and Mordred \cite{Moriwaki.2018}. We carried out this computation for every molecule in the PDBBind test set (\autoref{fig:single-rep-interpret}; scatter plots). We found that some channels were significantly correlated with more than one property (\autoref{tab:interpretability-r-values}, $p$-values were at least 10$^{-40}$; we did not correct for multiple hypothesis testing as there were only 64 channels), indicating that \textsc{QuickBind} has in fact learned physicochemical characteristics of protein-ligand binding. Selecting the most strongly correlated property per channel, we found the number of H-bond acceptors and donors, the total hydrophobic surface area, and the number of rotatatable bonds to be the strongest features. We include additional results in section~\ref{sec:add-interpret}.

\subsection{Case studies}
To better understand \textsc{QuickBind}'s utility in real-world deployments, we predicted the bound poses of new ligands for five proteins in the PDBBind test set, which we selected based on their extensive experimental characterization and clinical significance (UniProt IDs B1MDI3, P56817, P17931, Q8ULI9, and P01116, \autoref{tab:vs-targets}). B1MDI3 is a tRNA guanine-methyltransferase, P56817 is BACE1, a beta-secretase relevant to the development of Alzheimer's disease \cite{bace}, P17931 is galectin-3, a galactose-specific lectin involved in cancer \cite{galectin}, Q8ULI9 is the human immunodeficiency virus 1 (HIV-1) protease, and P01116 is the GTPase K-Ras, a key cancer target. We used the protein crystal structures of the three lowest affinity binders from the PDBBind test set as input, and predicted the binding affinities and complex structures of all other compounds in the PDBBind test set. We treated all compounds not explicitly crystallized with the five target proteins as decoys (Figures~\ref{fig:ba1}-\ref{fig:ba3}, \autoref{tab:vs-targets}). This likely resulted in some binders being mislabeled as non-binders.

We found \textsc{QuickBind} capable of discerning ligand binding characteristics across diverse ligand scaffolds and protein targets. Specifically, \textsc{QuickBind} distinguished binders from non-binders across various targets, including BACE1 ($p$-values of one-sided Wilcoxon rank sum tests of 0.0483, 0.0196, 0.0229), galectin-3 (0.0333, 0.0383, 0.0167), and K-Ras (< 0.00005, 0.0001, 0.0105), despite BACE1 and K-Ras binders having low average Tanimoto similarity. For BACE1 and galectin-3, \textsc{QuickBind} also accurately predicted poses with a ligand RMSD consistently below 2\AA\ for the majority of ligands (success rates of 73\% and 86\%, respectively), based on our comparisons with PDBBind test set structures. For the HIV-1 protease, while the predicted binding affinities did not significantly differentiate binders from non-binders, \textsc{QuickBind} still excelled at predicting pose structures with high accuracy (all predictions below 2\AA, \autoref{fig:crossdocking_preds}).

Our analyses also shed light on the limitations of \textsc{QuickBind}. It struggled when confronted with proteins whose conformation changes dramatically when binding target ligands versus low affinity ligands used as templates. For instance, KRAS structures had an average backbone RMSD of 7.7\AA\ between input and true conformations, which resulted in a notable proportion of poses exhibiting high ligand RMSDs (only 17\% of predictions fell below 5\AA). For the wholly unseen tRNA guanine-methyltransferase, \textsc{QuickBind} struggled to predict significantly higher binding affinities for binders vs. non-binders and only predicted moderately accurate poses with a ligand RMSD below 5\AA\ in 2\% of cases.

\section{Discussion}
\textsc{QuickBind} is an ML model for blind molecular docking that optimizes runtime speed while retaining competitive pose prediction accuracy, providing a compelling option for high-throughput virtual screening. It furthermore captures physicochemical ligand properties known to influence molecular docking. Interpretability informs what aspects of the underlying physics a model has captured and may, if sufficiently well-developed, help guide the design of drug compounds. These considerations are important given the large costs involved in identifying and prioritizing compounds for experimental testing and optimization.

Our augmentation of \textsc{QuickBind} with affinity prediction capabilities make it suitable for identifying new ligands, which we showcase using multiple highly relevant drug targets. Except for \textsc{TANKBind}, existing blind docking and co-folding models do not predict binding affinities, and are therefore not applicable for screening applications. The combination of ligand pose and interaction strength can help in optimizing the potency and selectivity of drug candidates, although it remains to be seen how sensitive \textsc{QuickBind} is to minor structural changes, which can have profound effects on binding (so-called activity cliffs).

An added advantage of \textsc{QuickBind}'s formulation of binding affinity prediction is its lack of reliance on an experimental co-complex structure (unlike \textsc{TANKBind}), due to having been trained separately from the main docking model. Embeddings from any protein-ligand pair can be used, whether the structure is predicted or experimentally-derived. This means that \textsc{QuickBind}'s affinity module can be trained on BindingDB \cite{BindingDB}, a binding affinities database comprising millions of protein-ligand pairs, orders of magnitude more than structural complexes. Users can also rapidly finetune a binding affinity model for their own target proteins.

Evaluation using the PB Benchmark showed that \textsc{QuickBind} struggles to generate physically and chemically valid poses, but that many can be recovered by force field-based energy minimization. While energy minimization can increase model runtime, the binding affinity module can be used to first select a subset of promising compounds, whose poses can then be energy minimized.

\section{Outlook}
A major shortcoming of \textsc{QuickBind} and other ML-based molecular docking methods is their dependence on rigid re-docking, a task on which they are both trained and evaluated. In rigid re-docking, methods are provided with the \textit{holo} protein structure that was originally co-crystallized with the query ligand, and the protein structure is not predicted. This does not reflect many real-world uses of molecular docking in which users only have access to the \textit{apo} (unbound) structure or a \textit{holo} structure co-crystallized with another ligand. Although \textsc{QuickBind} showed promise at cross-docking in the case studies we conducted, it did so only as long as the input protein did not deviate too much from the true structure. Ideally, models should be trained and evaluated in a flexible cross-docking setting, and in fact ML-based blind, flexible docking methods \cite{Lu.2024} have recently emerged. Future versions of \textsc{QuickBind} can be adapted into a flexible docking method capable of using \textit{apo}, \textit{holo}, or even predicted protein structures as input by updating protein residue frames as well as ligand frames. Alternatively, side chains can be made flexible by updating the rotational component of residue frames and side chain torsion angles.

Conceptually, one of our goals for \textsc{QuickBind} was to investigate how the \textsc{AF2} architecture may be adapted to the task of docking and co-folding. While other docking \cite{TANKBind, E3Bind} and co-folding methods \cite{RFAA, NeuralPLexer, UMol, Nakata.2023} previously demonstrated that ideas and components from \textsc{AF2} can be used in protein-ligand pose prediction, \textsc{QuickBind} uses essentially the entirety of \textsc{AF2}. This has provided multiple insights. First, it suggests that \textsc{QuickBind} could further benefit from \textsc{AF2}’s native confidence estimates and recycling, which were found to be important for \textsc{AF2}'s success. \textsc{QuickBind} could also benefit from more complicated, \textsc{AF2}-inspired loss functions, for example the structural violation loss which would likely lead to more PB-valid predictions. Second, \textsc{QuickBind} provides a rough estimate of how well an \textsc{AF2}-like model can perform molecular docking. The fact that it does not achieve state-of-the-art performance suggests that certain aspects of \textsc{AF2} are not ideally suited for this task and anticipates some of the changes introduced in \textsc{AF3}, including a minimized MSA module and atomistic reasoning over ligands. Including further innovations from \textsc{AF3} will likely result in an improved molecular docking tool (section~\ref{sec:af3-comp}). In turn, our findings also have implications for \textsc{AF3}. For instance, since \textsc{QuickBind}'s single representation captures physicochemical features of the ligand, it is likely that \textsc{AF3}'s single representation is highly information-dense as well, and may therefore be useful for tasks beyond pose prediction.

\begin{ack}
    We would like to thank Psivant Therapeutics for providing computational resources. W.T. acknowledges financial support from the Max Planck School Matter to Life. S.C.K. is supported by NIH grant R35GM150546.
\end{ack}

\newpage

\bibliographystyle{naturemag}
\bibliography{references}

\newpage

\appendix

\renewcommand\thefigure{SI.\arabic{figure}}    
\setcounter{figure}{0}

\renewcommand\thetable{SI.\arabic{table}}    
\setcounter{table}{0}

\renewcommand\thesection{SI.\arabic{section}}    
\setcounter{section}{0} 

\section*{Supplementary Information}
\section{Algorithms} \label{sec:algos}
\begin{algorithm}[htpb]
	\caption[QuickBind architecture.]{\textsc{QuickBind} architecture. Module names correspond to the names of the algorithms in the supplementary information of \cite{Jumper.2021} and the same notation is used. The Evoformer Stack does not contain the MSAColumnAttention module. New modules are highlighted in blue. Concatenation (concat) and deconcatenation (deconcat) happens along the sequence and atom dimension, unless otherwise stated. $\left\{r_i\right\}$, protein features. $\left\{l_i\right\}$, ligand features. $\left\{\overrightarrow{x}_{\mathrm{C}_\alpha} \right\},\ \left\{\overrightarrow{x}_{\mathrm{C}} \right\},\ \left\{\overrightarrow{x}_{\mathrm{N}} \right\}$, coordinates of C$_\alpha$, C, and N atoms. $\left\{\overrightarrow{t}_i^\mathrm{lig}\right\}$, initial ligand coordinates. $\left\{f_i^\mathrm{res\_index}\right\}$, indices of amino acid residues. $\left\{f_{ij}^\mathrm{adj}\right\}$, ligand adjacency matrix.}
	\label{alg:QuickBind}
	\begin{algorithmic}[1]
		\Function{QuickBind}{$\left\{r_i\right\},\ \left\{l_i\right\},\ \left\{\overrightarrow{x}_{\mathrm{C}_\alpha} \right\},\ \left\{\overrightarrow{x}_{\mathrm{C}} \right\},\ \left\{\overrightarrow{x}_{\mathrm{N}} \right\},\ \left\{\overrightarrow{t}_i^\mathrm{lig}\right\},\ \left\{f_i^\mathrm{res\_index}\right\},\ \left\{f_{ij}^\mathrm{adj}\right\}$}
            \State $\left\{\overrightarrow{X}_\mathrm{C}^\mathrm{pseudo}\right\}, \left\{\overrightarrow{X}_\mathrm{N}^\mathrm{pseudo}\right\} \gets \mathrm{\textcolor{NavyBlue}{getadjacentatoms}}\left(\left\{\overrightarrow{t}_i^\mathrm{lig}\right\} \right)$ \Comment{as described in \hyperref[sec:architecture]{Methods}}
            \State $T_i^\mathrm{lig} \gets \mathrm{rigidFrom3Points} \left(\overrightarrow{t}_i^\mathrm{lig}, \overrightarrow{x}_{\mathrm{C}}^\mathrm{pseudo}, \overrightarrow{x}_{\mathrm{N}}^\mathrm{pseudo}\right)$
			\State $T_i^\mathrm{prot} \gets \mathrm{rigidFrom3Points} \left(\overrightarrow{x}_{\mathrm{C}_\alpha}, \overrightarrow{x}_{\mathrm{C}}, \overrightarrow{x}_{\mathrm{N}}\right)$
			\State $\left\{T_i \right\} \gets \mathrm{concat}\left(\left\{T_i^\mathrm{prot} \right\},\ \left\{T_i^\mathrm{lig} \right\}\right)$
			\State $\left\{s_i\right\},\ \left\{z_{ij}\right\} \gets \hyperref[alg:inputembedder]{\mathrm{\textcolor{NavyBlue}{InputEmbedder}}}\left(\left\{r_i\right\},\ \left\{l_i\right\},\ \left\{\overrightarrow{x}_{\mathrm{C}_\alpha} \right\},\  \left\{\overrightarrow{t}_i^\mathrm{lig}\right\},\ \left\{f_i^\mathrm{res\_index}\right\},\ \left\{f_{ij}^\mathrm{adj}\right\}\right)$ 
			\State \LeftComment{Evoformer}
			\State $\left\{s_i\right\},\ \left\{z_{ij}\right\} \gets \mathrm{EvoformerStack}\left(\left\{s_i\right\},\ \left\{z_{ij}\right\}\right)$ 
            		\State \LeftComment {Structure Module}
			\State $s_i \gets \mathrm{LayerNorm}\left(s_i\right)$
			\State $z_{ij} \gets \mathrm{LayerNorm}\left(z_{ij}\right)$
			\State $s_i \gets \mathrm{Linear}\left(s_i\right)$ 
			\ForAll{$l\in\left[1, \dots, N_\mathrm{Struct}\right]$}
				\State $\left\{s_i\right\} \mathrel{+}= \mathrm{InvariantPointAttention}\left(\left\{s_i\right\},\ \left\{z_{ij}\right\}, \left\{T_i\right\} \right)$
				\State $s_i \gets \mathrm{LayerNorm}\left(\mathrm{Dropout}_{0.1}\left(s_i\right) \right)$
				\State $s_i \gets s_i + \mathrm{Linear}\left( \mathrm{relu} \left( \mathrm{Linear} \left( \mathrm{relu} \left( \mathrm{Linear} \left(s_i\right)\right)\right)\right)\right)$
				\State $s_i \gets \mathrm{LayerNorm}\left(\mathrm{Dropout}_{0.1}\left(s_i\right) \right)$
 			\State $\left\{T_i^\mathrm{prot} \right\},\ \left\{T_i^\mathrm{lig} \right\} \gets \mathrm{deconcat}\left(\left\{T_i \right\}\right)$
                \State $T_i^\mathrm{lig} \gets T_i^\mathrm{lig} \circ \hyperref[alg:backboneupdate]{\mathrm{\textcolor{NavyBlue}{BackboneUpdate}}}\left(s_i\right)$ \Comment{Use predicted quaternion and}
			  \State $\left\{T_i \right\} \gets \mathrm{concat}\left(\left\{T_i^\mathrm{prot} \right\},\ \left\{T_i^\mathrm{lig} \right\}\right)$ \Comment{translation to update ligand frames}
			\EndFor
   		\State $\left\{\mathbf{R}_i^\mathrm{lig}\right\}, \left\{\overrightarrow{t}_i^\mathrm{lig}\right\} = \left\{T_i^\mathrm{lig}\right\} $
			\State \textbf{return} $\left\{\overrightarrow{t}_i^\mathrm{lig}\right\}$
		\EndFunction
    \end{algorithmic}
\end{algorithm}

\begin{algorithm}
	\caption[Algorithm for the generation of input embeddings.]{Algorithm for the generation of input embeddings.}
	\begin{algorithmic}[1]
		\Function{InputEmbedder}{$\left\{r_i\right\},\ \left\{l_i\right\},\ \left\{\overrightarrow{x}_{\mathrm{C}_\alpha} \right\},\ \left\{\overrightarrow{t}_i^\mathrm{lig}\right\},\ \left\{f_i^\mathrm{res\_index}\right\},\ \left\{f_{ij}^\mathrm{adj}\right\}$}
			\State $r_i \gets \mathrm{Linear}\left(r_i\right)$ 
			\State $l_i \gets \mathrm{Linear}\left(l_i\right)$
			\State $\left\{s_i\right\} \gets \mathrm{concat}\left(\left\{r_i\right\},\ \left\{l_i\right\}\right)$ 
			\State $a_i \gets \mathrm{Linear}\left(s_i\right)$ 
			\State $b_i \gets \mathrm{Linear}\left(s_i\right)$ 
			\State $z_{ij} \gets a_i + b_j$
			\State $\left\{p_{ij}\right\} \gets \mathrm{relpos}\left(\left\{f_i^\mathrm{res\_index}\right\}\right)$
			\State $q_{ij} \gets \mathrm{Linear}\left(f_{ij}^\mathrm{adj}\right)$
			\State $\left\{\omega_{ij} \right\} \gets \mathrm{blockdiag}\left(\left\{p_{ij}\right\},\ \left\{q_{ij}\right\} \right)$ \Comment{combine into a blockdiagonal tensor}
			\State $d_{ij}^\mathrm{prot} \gets \mathrm{Linear}\left(\left\|\overrightarrow{x}_{\mathrm{C}_\alpha}^i-\overrightarrow{x}_{\mathrm{C}_\alpha}^j\right\|_2 \right)$ 
   		\State $d_{ij}^\mathrm{lig} \gets \mathrm{Linear}\left(\left\|\overrightarrow{t}^\mathrm{lig}_i-\overrightarrow{t}_j^\mathrm{lig}\right\|_2 \right)$
     	\State $\left\{d_{ij} \right\} \gets \mathrm{blockdiag}\left(\left\{d_{ij}^\mathrm{prot}\right\},\ \left\{d_{ij}^\mathrm{lig}\right\} \right)$ \Comment{combine into a blockdiagonal tensor}
			\State $z_{ij} = z_{ij} + \omega_{ij} + d_{ij}$
			\State $s_i \gets \mathrm{Linear}\left(s_i\right)$
			\State \textbf{return} $\left\{s_i\right\},\ \left\{z_{ij}\right\}$
		\EndFunction
    \end{algorithmic}
		\label{alg:inputembedder}
\end{algorithm}

\begin{algorithm}
	\caption[Algorithm for the backbone update step.]{Algorithm for updating the ligand frames.}
	\begin{algorithmic}[1]
		\Function{BackboneUpdate}{$\left\{s_i\right\}$}
			\State $\left\{s_i^\mathrm{prot}\right\}, \left\{s_i^\mathrm{lig}\right\} \gets \mathrm{deconcat} \left(\left\{s_i\right\}\right)$
            \State $m_i \gets \mathrm{MultiHeadAttention} \left(q=\left\{s_i^\mathrm{lig}\right\}, k=\left\{s_i^\mathrm{prot}\right\}, v=\left\{s_i^\mathrm{prot}\right\} \right)$
            \State $s_i^\mathrm{lig} \mathrel{+}= m_i$
			\State $b_i, c_i, d_i, \overrightarrow{t}_i \gets \mathrm{Linear} \left(s_i^\mathrm{lig} \right)$ \Comment{quaternion and translation}
			\State \textbf{return} $b_i, c_i, d_i, \overrightarrow{t}_i$
		\EndFunction
    \end{algorithmic}
		\label{alg:backboneupdate}
\end{algorithm}

\begin{algorithm}
	\caption[Dummy atom construction.]{Algorithm for getting the coordinates of a dummy atom when an atom has only one neighbor. Find the vector with the same $x$ and $y$ coordinates as the bond vector between the atom under question and its one neighbor, such that the dot product of the two vanishes, and subtract that vector from the coordinates of the given atom. $\overrightarrow{x}_\mathrm{C}$, coordinates of atom under consideration, $\overrightarrow{x}_\mathrm{N}$, coordinates of its neighbor.}
	\begin{algorithmic}[1]
		\Function{GetDummyAtomCoords}{$\overrightarrow{x}_\mathrm{C}$, $\overrightarrow{x}_\mathrm{N}$}
			\State $\overrightarrow{b} \gets \overrightarrow{x}_\mathrm{C} - \overrightarrow{x}_\mathrm{N}$
			\State $x_b, y_b, z_b \gets \overrightarrow{b}$ \Comment{get $x$, $y$, $z$ coordinates}
            \State $z_b' \gets -\dfrac{x_b^2 + y_b^2}{z_b}$ 
			\State $\overrightarrow{b}' \gets \left(\begin{matrix}x_b\\y_b\\ z_b'\end{matrix}\right)$ \Comment{$\overrightarrow{b} \cdot \overrightarrow{b}'=0$}
            \State $\overrightarrow{x}_\mathrm{N'} \gets \overrightarrow{x}_\mathrm{C} - \overrightarrow{b}'$
			\State \textbf{return} $\overrightarrow{x}_\mathrm{N'}$
		\EndFunction
    \end{algorithmic}
		\label{alg:dummyatom}
\end{algorithm}

\begin{algorithm}
	\caption[Binding affinity prediction.]{Binding affinity prediction model.}
	\begin{algorithmic}[1]
		\Function{BindingAffinityPredictor}{$\left\{s_i\right\}$}
			\State $s_i \gets \mathrm{LayerNorm}\left(s_i\right)$
			\State $s_i \gets \mathrm{Linear}\left(\mathrm{silu}\left(\mathrm{Linear}\left(s_i\right)\right)\right)$
            \State $a \gets \mathrm{mean}\left(\left\{s_i\right\}\right)$ \Comment{along sequence dimension}
			\State $a \gets \mathrm{Linear}\left(\mathrm{relu}\left(\mathrm{Linear}\left(\mathrm{relu}\left(\mathrm{Linear}\left(a\right)\right)\right)\right)\right)$ \Comment{$\mathbb{R}^{64}\rightarrow\mathbb{R}^{32}\rightarrow\mathbb{R}$}
        \State \textbf{return} $a$
		\EndFunction
    \end{algorithmic}
		\label{alg:binding-affinity-prediction}
\end{algorithm}

\newpage

\section{Protein and ligand features} \label{sec:feats}
Residue types are one-hot encoded. Ligand atoms have the following features: atomic number (H, C, N, O, F, P, S, Cl, Br, I, other), chirality, degree (1 through 4, or other), formal charge (-1, 0, 1, or other), number of connected H atoms (0 through 3, or other), hybridization, presence in a ring, and presence in an aromatic ring. Early versions included atomic numbers 1 through 119, degrees up to 10, formal charges of -5 through 5, implicit valence, number of connected hydrogens of up to 8, number of radical electrons, hybridization, presence in an aromatic ring, the number of rings it is in, presence in a ring of size 3, 4, 5, 6, 7, or 8, similar to ref. \cite{EquiBind}. These additional features did not improve performance and so were omitted. Ligand coordinates are initialized using a random RDKit conformer \cite{rdkit}.

\section{Cropping}\label{sec:cropping}
To reduce \textsc{QuickBind}'s memory footprint during training, input protein sequences were cropped to 512 or 256 residues for models without and with the Evoformer module, respectively. Since model performance drops noticeably for shorter crop sizes, the final model was finetuned with a crop size of 512 residues. At inference time, the full protein sequence is used. Depending on available GPU memory, inference might therefore have to be run on CPUs. Using our resources, we had to restrict inference on GPUs to proteins shorter than 2,000 residues. For multi-chain proteins, sequences were concatenated in the order in which they appear in the PDB file.

We tested different cropping strategies: random contiguous cropping; setting the residue whose C$_\alpha$ atom is closest to any ligand atom as the midpoint of the contiguous cropped fragment (binding site cropping, BSC); selecting the $x$ residues closest to any ligand atom (spatial cropping), and a setting in which the protein was cropped randomly or spatially with a probability of 0.5.

\section{Hyperparameter screening}\label{sec:hyperparamsscreen}
Because extensive hyperparameter screening of the full \textsc{QuickBind} model would have been too computationally expensive, we optimized many hyperparameters and architectural choices using two smaller variants, both trained without ligand frames and still updating H atom positions:
\begin{itemize}
    \item \textsc{QuickBind-S}, which lacks the Evoformer module and contains just four or eight Structure module blocks with unshared weights, trained without batching.
    \item \textsc{QuickBind-M}, which lacks Triangle Attention in the Evoformer stack (its most expensive module), trained with a batch size of 12 or 16.
\end{itemize}
The main results of the hyperparameter screening are summarized in \autoref{tab:hyperparameter-screening}. Furthermore, several ways to generate the ligand coordinate updates from the final single representation were tested. In general, the final single representation is separated into a protein and a ligand single representation. Then, either:
\begin{itemize}
    \item only the ligand single representation is passed through a linear layer to produce the coordinate updates, or
    \item the protein single representation is summed along the sequence dimension or the corresponding mean is taken, and this pooled protein representation is concatenated with the ligand single representation before passing it through the linear layer, or
    \item the outer product of the protein and ligand single representations is summed along the sequence dimension or the corresponding mean is taken, and then passed through the linear layer, or 
    \item the output of an attention layer with the query vectors coming from the ligand and key and value vectors coming from the protein are concatenated with the ligand single representation and passed through the final linear layer.
\end{itemize}
The last approach led to the best results.

We also tested scaling the coordinate update by a factor of 10, as is done in \textsc{AF2}, but did not find this to improve model performance. Furthermore, we briefly experimented with first applying a global rototranslation of the ligand coordinates, and then either finetuning all ligand coordinates or just changing the torsion angles of rotatable bonds, but did not find this to lead to better results. Using a gated variant of the IPA module \cite{gatedIPA} improved model performance compared to the standard IPA module. In addition, we also tested two ideas from \textsc{AlphaFold-Multimer} (\textsc{AF-Multimer}) \cite{AFMultimer}, moving the outer product mean to the beginning of the Evoformer block and the multimer version of the relative positional encoding, but neither improved model performance.

\begin{table}[htpb]
    \caption{Results of hyperparameter search. Accepted configurations are indicated by \cmark or underlined, rejected configurations are indicated by \xmark. $\mathcal{L}_\mathrm{MSE}$ - Mean squared error loss. $\mathcal{L}_\mathrm{centroid}$ - Centroid loss. $\mathcal{L}_\mathrm{Kabsch}$ - Kabsch loss. $\mathcal{L}_\mathrm{FAPE}$ - FAPE loss. $\mathcal{L}_\mathrm{FAPE}$ was tested with a $\mathcal{L}_\mathrm{FAPE}^\mathrm{prot-lig}$ clamped at 10 \AA\ and without clamping. $\mathcal{L}_\mathrm{FAPE}^\mathrm{aux}$ - intermediate FAPE losses acting on the outputs of every Structure module block. $\mathcal{L}_\mathrm{dist}^\mathrm{lig-lig}$ - ligand distogram loss head, a cross-entropy loss that acts on a symmetrized version of the ligand pair representation, similar to \textsc{AF2}, with 42 distance bins between 1 and 5 \AA. $\mathcal{L}_\mathrm{dist}^\mathrm{prot-lig}$ - protein-ligand distogram loss head using the symmetrized off-diagonal parts of the pair representation, using the same bins as in \textsc{AF2}. $\mathcal{L}_\mathrm{torsion}$ - torsion angle loss. Black hole initialization refers to collapsing all ligand atoms at the origin, as is done in \textsc{AF2}.}
    \centering
    \begin{tabular}{lc} \toprule
        \textbf{Input embeddings} & \\
        Pairwise distances & \cmark \\
        Radial basis projection of pairwise distances & \xmark \\
        \textsc{AF2} relative positional encoding & \cmark \\
        \textsc{AF2-Multimer} relative positional encoding & \xmark \\
        Adjacency matrix &  \cmark \\
        ... w/ one-hot encoded bond types &  \xmark \\
        ... w/ topological distance & \xmark \\ \midrule
        \textbf{Loss function} \\
        \multicolumn{2}{l}{$\mathcal{L}_\mathrm{MSE}$, $\mathcal{L}_\mathrm{centroid}$, \underline{$\mathcal{L}_\mathrm{Kabsch}$}, \underline{$\mathcal{L}_\mathrm{FAPE}$} (clamped and \underline{unclamped})}, \underline{$\mathcal{L}_\mathrm{FAPE}^\mathrm{aux}$}, $\mathcal{L}_\mathrm{dist}^\mathrm{lig-lig}$, $\mathcal{L}_\mathrm{dist}^\mathrm{prot-lig}$, $\mathcal{L}_\mathrm{torsion}$ \\ \midrule
        \textbf{Cropping} \\
        Random  & \xmark \\
        Binding site cropping &  \cmark \\
        Spatial cropping  & \xmark \\
        Random and spatial cropping & \xmark \\ \midrule
        \textbf{Ligand frames} \\
        Keeping the rotation matrix fixed & \xmark \\
        Updating the rotation matrix & \cmark \\ \midrule
        \textbf{Ligand initialisation} \\
        At origin & \cmark \\
        Black hole initialisation & \xmark \\
        Randomly translated and rotated & \xmark \\ \midrule
        \textbf{Structure module} \\
        Number of blocks & 4, \underline{8} \\ \midrule
        \textbf{Evoformer} \\
        Number of blocks & 8, \underline{12} \\
        Number of MSA attention heads & \underline{8}, 12 \\
        \bottomrule
    \end{tabular}
    \label{tab:hyperparameter-screening}
\end{table}

\section{Training details} \label{sec:train-details}
\textsc{QuickBind} was implemented using PyTorch \cite{PyTorch}, PyTorch Lightning \cite{PyTorchLightning}, OpenFold \cite{OpenFold}, and the RDKit \cite{rdkit}. Models were trained using the AdamW optimizer \cite{AdamW} with a learning rate and a weight decay coefficient of $10^{-4}$, early stopping with a patience of 50 epochs, and a batch size of 16. The binding affinity prediction model was trained using the Adam optimizer \cite{Adam} with a learning rate of 0.01, early stopping with a patience of 50 epochs, and a batch size of 64 using a mean squared error (MSE) loss. The model weights with the best performance on the validation set were chosen for evaluation on the test set. Training the final \textsc{QuickBind} model took several weeks on eight NVIDIA A40 GPUs, but \textsc{QuickBind-S} and \textsc{QuickBind-M} variants were trained in two weeks or less. The final model was trained with 5 different seeds. Some replicas got stuck in local minima (success rates of 0.0\%, 2.5\%, 12.1\%, 15.4\%, 23.4\%), and only the best-performing model was finetuned with a crop size of 512. It is plausible that models that got stuck in local minima would have reached a similar performance to the final model after the finetuning stage, but we did not test this because the training time when including triangle attention scales very unfavorably with sequence length. Force-field minimization was performed as described in ref. \cite{Buttenschoen.2023} using a script kindly provided by one of the authors. Visualizations were generated using NGLview \cite{nglview}.

\clearpage

\section{Exemplary predictions}
\begin{figure}[htpb]
     \centering
     \begin{subfigure}[b]{0.3\linewidth}
         \centering
         \includegraphics[width=\linewidth]{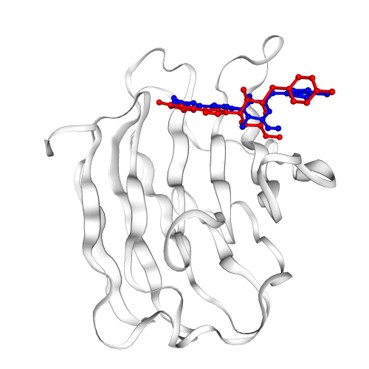}
         \caption{1.8\AA.}
         \label{fig:example-a}
     \end{subfigure}
     \hfill
     \begin{subfigure}[b]{0.3\linewidth}
         \centering
         \includegraphics[width=\linewidth]{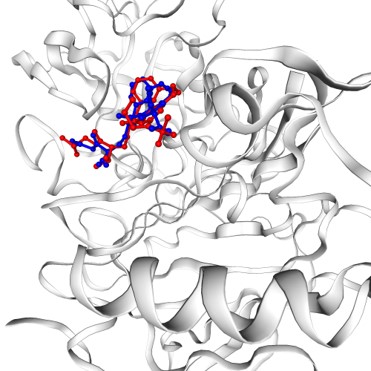}
         \caption{1.1\AA.}
     \end{subfigure}
     \hfill
     \begin{subfigure}[b]{0.3\linewidth}
         \centering
         \includegraphics[width=\linewidth]{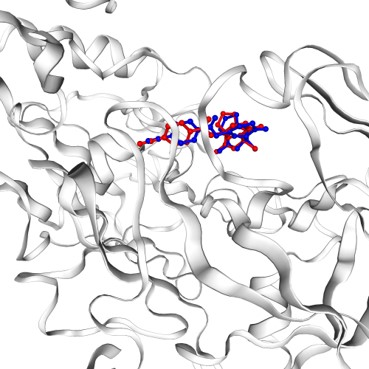}
         \caption{0.7\AA.}
     \end{subfigure}
        \caption{Three examples of \textsc{QuickBind} predictions and their RMSDs, randomly chosen from 100 lowest-RMSD predictions on the PDBBind test set. The ground-truth ligand is shown in red, the \textsc{QuickBind} prediction is shown in blue. }
        \label{fig:examples}
\end{figure}

\newpage
 
\section{PB failure modes}

\begin{figure}[htpb]
    \centering
     \begin{subfigure}[b]{\linewidth}
         \centering
         \includegraphics[width=.6\linewidth]{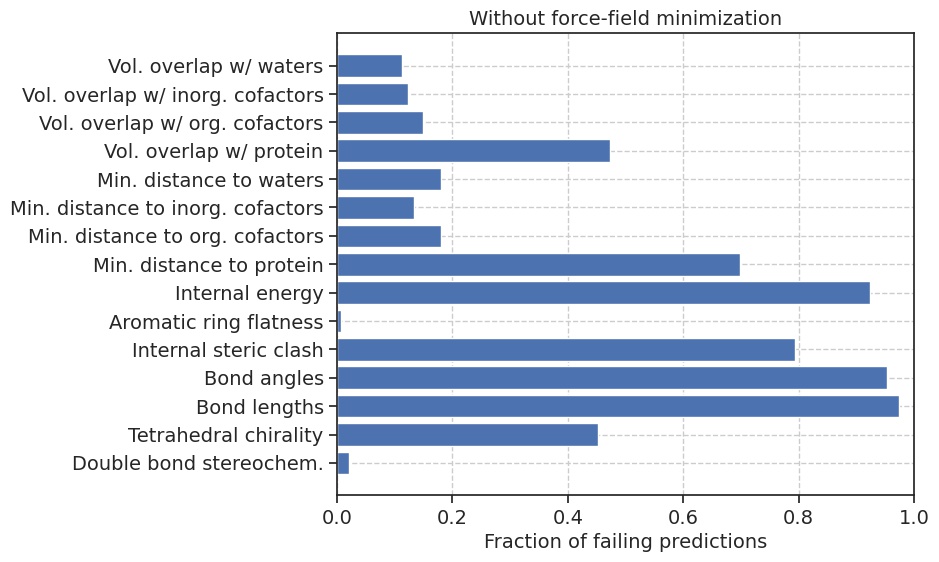}
     \end{subfigure}
      \begin{subfigure}[b]{.6\linewidth}
         \centering
         \includegraphics[width=\linewidth]{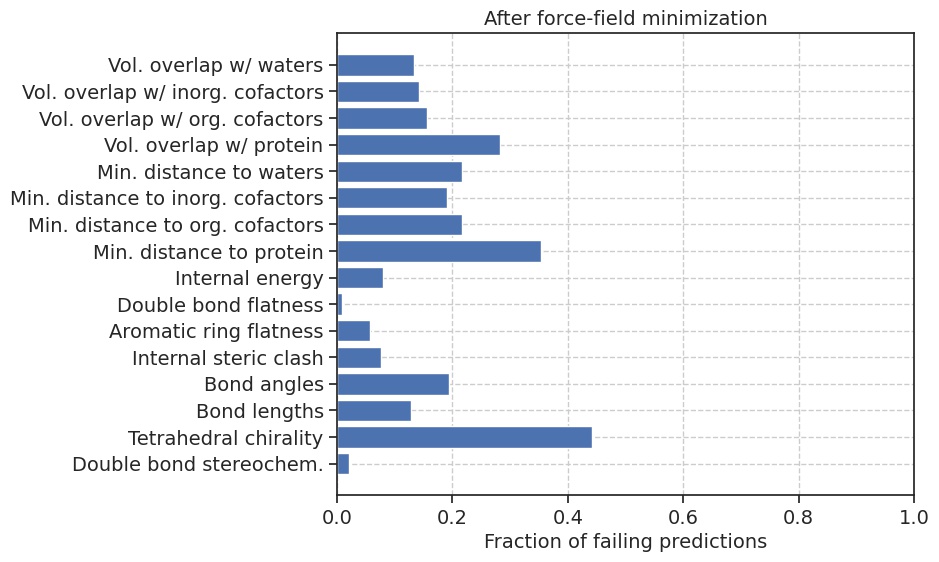}
     \end{subfigure}
    \caption{Fraction of all \textsc{QuickBind} predictions on the PB test set that fail PB tests, before and after force field minimization.}
    \label{fig:fail-modes}
\end{figure}

\newpage

\section{Correlation with physicochemical features}

\begin{table}[htpb]
    \centering
    \caption{Correlation between physicochemical features of the ligand and mean channel values of the molecule representation. The magnitude of the Pearson's $R$ values for the octanol-water partition coefficient and number of aromatic rings was less than 0.6.}
    \begin{tabular}{cccc}
        \toprule
        Feature & Pearson's $R$ & $p$-value & Channel \\
        \midrule
        Total hydrophobic surface area & -0.72 & 10$^{-59}$ & 57 \\
        Molecular weight & 0.73 & 10$^{-62}$ & 27 \\
        Number of hydrogen bond acceptors & 0.63 & 10$^{-42}$ & 11 \\
        Number of hydrogen bond donors & 0.70 & 10$^{-54}$ & 31 \\
        Polar surface area & 0.71 &  10$^{-56}$ & 27 \\
        Number of rotatable bonds & 0.76 & 10$^{-70}$ & 27 \\
        \bottomrule
    \end{tabular}
    \label{tab:interpretability-r-values}
\end{table}

\newpage

\section{Additional interpretability studies} \label{sec:add-interpret}
\begin{figure}[htpb]
     \centering
     \begin{subfigure}[b]{\linewidth}
         \centering
         \includegraphics[width=\linewidth]{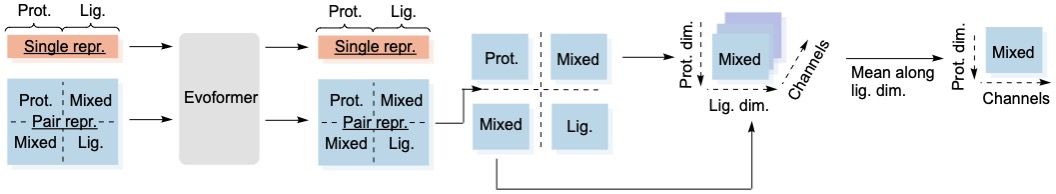}
     \end{subfigure}
     \begin{subfigure}[b]{0.3\linewidth}
         \centering
         \includegraphics[width=\linewidth]{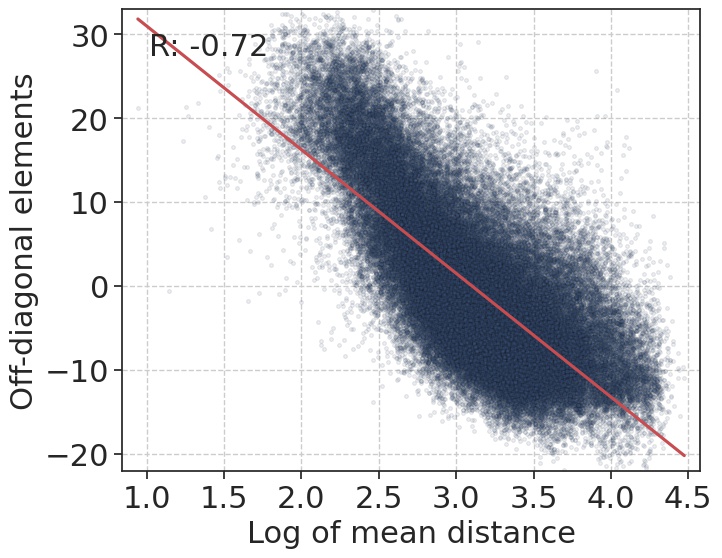}
     \end{subfigure}
        \caption{The off-diagonal elements of the pair representation contain information about the final ligand atom positions relative to the protein. The off-diagonal elements of the pair representation after the Evoformer block were symmetrized and the mean along the ligand dimension was taken. Some channel values correlate with the logarithm of the mean C$_\alpha$-ligand atom distance; for one particular channel the Pearson's R value is -0.72 at a $p$-value smaller than machine precision.}
        \label{fig:pair-rep-interpret}
\end{figure}

We wanted to understand how \textsc{QuickBind} obtains its initial guess of the docked ligand pose. The pair representation contains a protein and a ligand block, as well as mixed off-diagonal elements. Among other features, the protein and ligand blocks are constructed from the pairwise distances of the C$_\alpha$ and ligand atoms, respectively, but the off-diagonal elements do not contain any spatial information. We hypothesized that the model would use these off-diagonal elements for information about the interaction of the protein and the ligand, including an initial guess about their pairwise distances. We therefore took the off-diagonal elements of the pair representation after the Evoformer block, symmetrised them by transposing the lower off-diagonal block and computing the element-wise mean, and finally took the mean along the ligand dimension to obtain an $N_{\mathrm{C}_\alpha} \times c$ dimensional matrix, where $N_{\mathrm{C}_\alpha}$ is the number of C$_\alpha$ atoms and $c$ is the hidden channel dimension (\autoref{fig:pair-rep-interpret}). Indeed, we found that already the mean along the channel dimension is weakly correlated with the logarithm of the mean C$_\alpha$-ligand atom distance with a Pearson's R value of -0.53. This correlation is much stronger for some channels. In particular, there is a channel that correlates with the logarithm of the mean C$_\alpha$-ligand atom distance with a Pearson's R value of -0.72.

\newpage

\section{Binding affinity prediction}

\begin{table}[htpb]
    \centering
    \caption{Root-mean-square error (RMSE), Pearson correlation coefficient (PCC), Spearman's rank correlation coefficient (SRCC), and mean absolute error (MAE) of \textsc{QuickBind} and other methods for binding affinity prediction on the PDBBind test set, computed using the mean and standard deviation across three runs. All methods predict negative log-transformed binding affinities. Values for other methods are taken from ref. \cite{TANKBind}.}
    \begin{tabular}{lcccc}
        \toprule
        Method & RMSE $\downarrow$ & PCC $\uparrow$ & SRCC $\uparrow$ & MAE $\downarrow$ \\
        \midrule
        TransformerCPI \cite{Chen.2020} & $1.741 \pm 0.058$ & $0.576 \pm 0.022$ & $0.540 \pm 0.016$ & $1.404 \pm 0.040$ \\
        MONN \cite{Li.2020} & $1.438 \pm 0.027$ & $0.624 \pm 0.037$ & $0.589 \pm 0.011$ & $1.143 \pm 0.052$ \\
        PIGNet \cite{Moon.2022} & 2.64 & 0.51 & 0.49 & 2.1 \\
        IGN \cite{Jiang.2021a} & $1.433 \pm 0.028$ & $0.698 \pm 0.007$ & $0.641 \pm 0.014$ & $1.169 \pm 0.036$\\
        HOLOPROT \cite{Somnath.2021} & $1.546 \pm 0.065$ & $0.602 \pm 0.006$ & $0.571 \pm 0.018$ & $1.208 \pm 0.038$\\
        STAMPDPI \cite{Wang.2022} & 1.658 & 0.545 & 0.411 & 1.325 \\
        TANKBind \cite{TANKBind} & $1.346 \pm 0.007$ & $0.726 \pm 0.007$ & $0.703 \pm 0.017$ & $1.070 \pm 0.019$ \\
        \midrule
        QuickBind & $1.577 \pm 0.011$ & $0.548 \pm 0.025$ & $0.482 \pm 0.024$ & $1.292 \pm 0.008$\\
        \bottomrule
    \end{tabular}
    \label{tab:binding-affinity}
\end{table}

\newpage

\section{Virtual screening results}
\autoref{tab:vs-targets} summarizes important characteristics of the five proteins in the PDBBind test set with the highest number of complex structures in the PDBBind test set, as well as \textsc{QuickBind}'s cross-docking performance. In particular, it contains:
\begin{itemize}
    \item the number of complex structures in the PDBBind test set (\# Binders),
    \item the number of complex structures in the PDBBind train set (\# Train ex.),
    \item the average C$_\alpha$ RMSD between input and true protein structures (BB RMSD), 
    \item and the average Tanimoto similarity between the lowest, second-lowest, or third-lowest affinity binder and the remaining binders, calculated using extended-connectivity fingerprints \cite{Rogers.2010} with a radius of 3 (TS).
\end{itemize}
We evaluate \textsc{QuickBind}'s cross-docking performance using the fraction of predictions with a ligand RMSD below 2\AA\ (\% < 2\AA) and the fraction of predictions with a ligand RMSD below 5\AA\ (\% < 5\AA), after aligning the C$_\alpha$ atoms of the input and the true protein structure using the Kabsch algorithm. For this alignment and for calculating BB RMSD we only consider C$_\alpha$ atoms that were successfully extracted for both complexes. Where applicable, we provide the mean and standard deviation across the three runs with the crystal structures of the lowest, second-lowest, and third-lowest affinity binder. For all proteins, we tested if the predicted binding affinities of binders were higher than those of non-binders using one-sided Wilcoxon rank-sum tests.

\begin{table}[htpb]
    \centering
    \caption{Characteristics of the five proteins in the PDBBind test set with the highest number of complex structures and \textsc{QuickBind}'s cross-docking performance.}
    \begin{tabular}{ccccccc}
        \toprule
        UniProt ID & \# Binders & \# Train ex. & \% < 2 \AA\ & \% < 5 \AA\ & BB RMSD [\AA] & TS \\
        \midrule
        B1MDI3 & 19 & 0 & $0.0 \pm 0.0$ & $2 \pm 3$ & $0.36 \pm 0.09$ & $0.26 \pm 0.08$ \\
        P56817 & 16 & 308 & $73 \pm 0$ & $93 \pm 0$ & $0.96 \pm 0.04$ & $0.209 \pm 0.005$ \\
        P17931 & 15 & 22 & $86 \pm 0$ & $93 \pm 0$ & $0.21 \pm 0.07$ & $0.620 \pm 0.029$ \\
        Q8ULI9 & 14 & 3 & $100 \pm 0$ & $100 \pm 0$ & $0.15 \pm 0.01$ & $0.64 \pm 0.10$ \\
        P01116 & 13 & 8 & $0.0 \pm 0.0$ & $17 \pm 12$ & $7.7 \pm 2.7$ & $0.101 \pm 0.026$ \\
        \bottomrule
    \end{tabular}
    \label{tab:vs-targets}
\end{table}

\begin{figure}
    \centering
    \includegraphics[trim=0 350 0 0,clip,width=\linewidth]{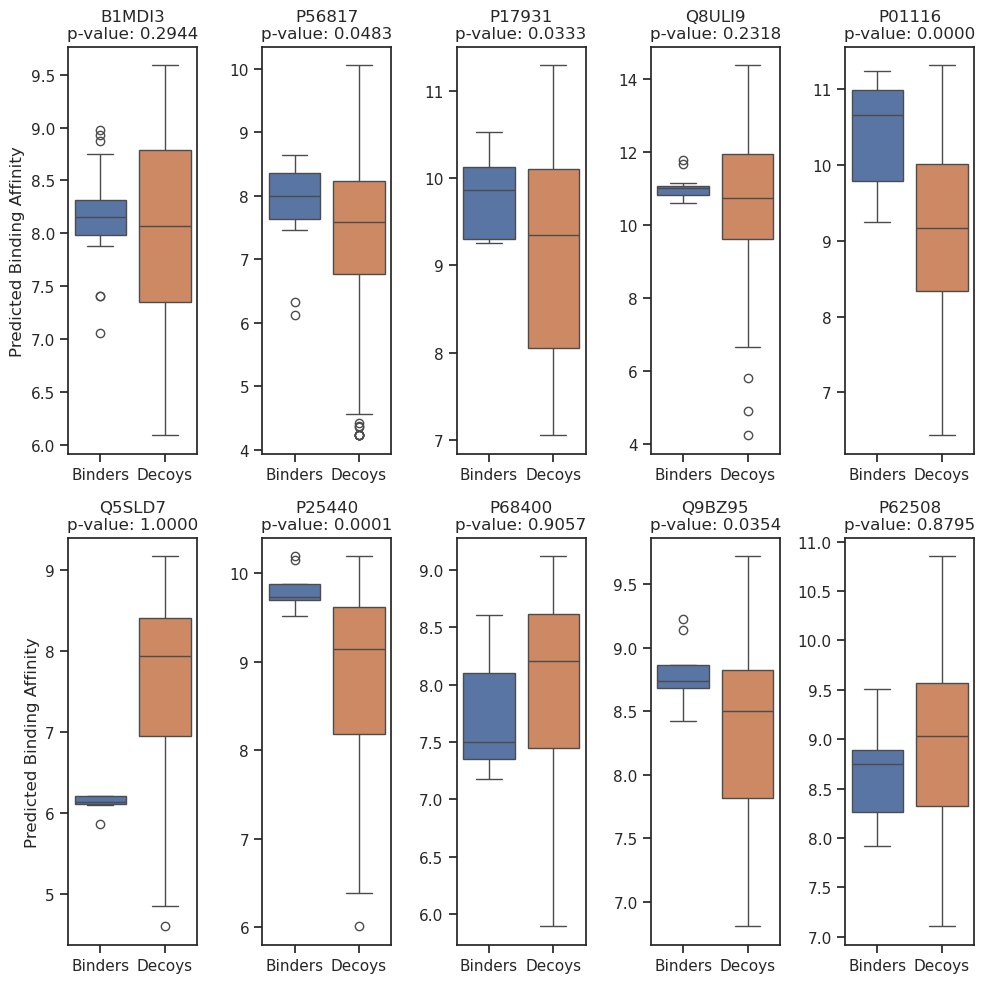}
    \caption{Predicted binding affinities of true binders and decoys for the five proteins in the PDBBind test set with the highest number of binders, using the protein structure of the lowest affinity binder.}
    \label{fig:ba1}
\end{figure}

\begin{figure}
    \centering
    \includegraphics[trim=0 350 0 0,clip,width=\linewidth]{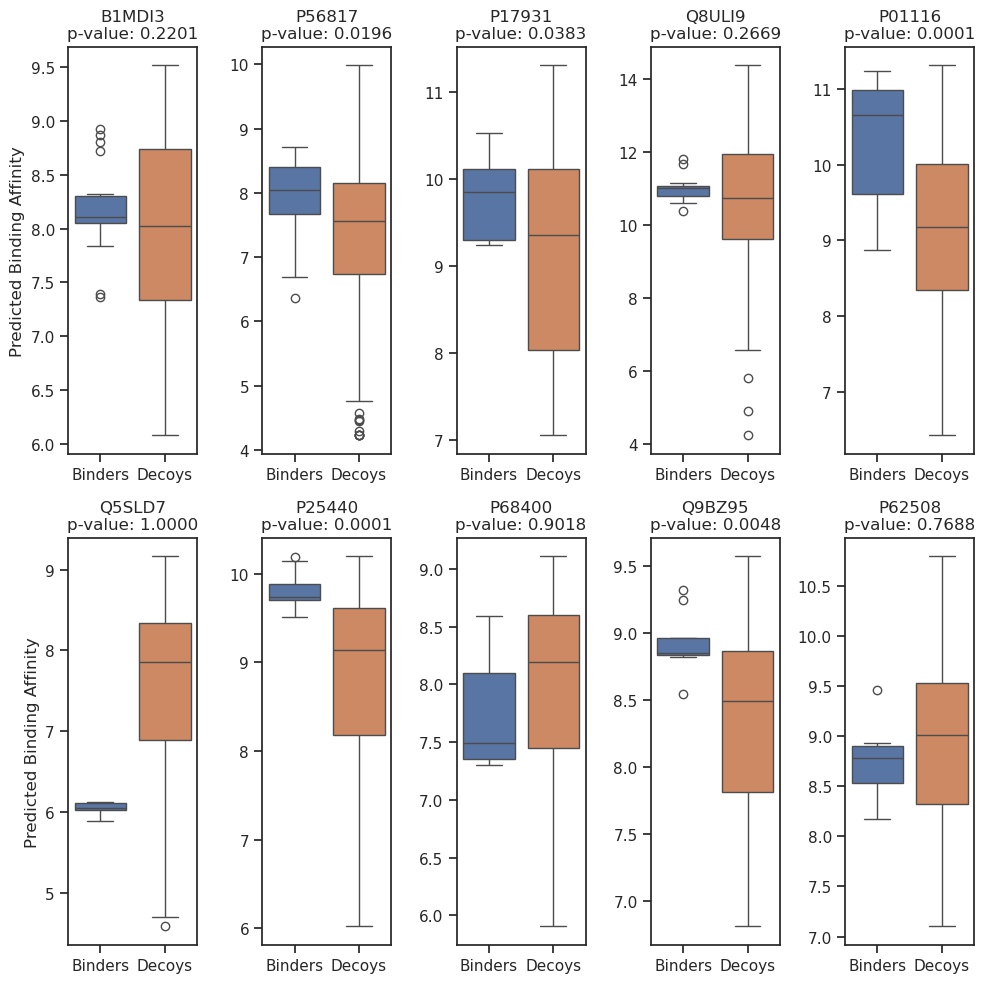}
    \caption{Predicted binding affinities of true binders and decoys for the five proteins in the PDBBind test set with the highest number of binders, using the protein structure of the second-lowest affinity binder.}
    \label{fig:ba2}
\end{figure}

\begin{figure}
    \centering
    \includegraphics[trim=0 350 0 0,clip,width=\linewidth]{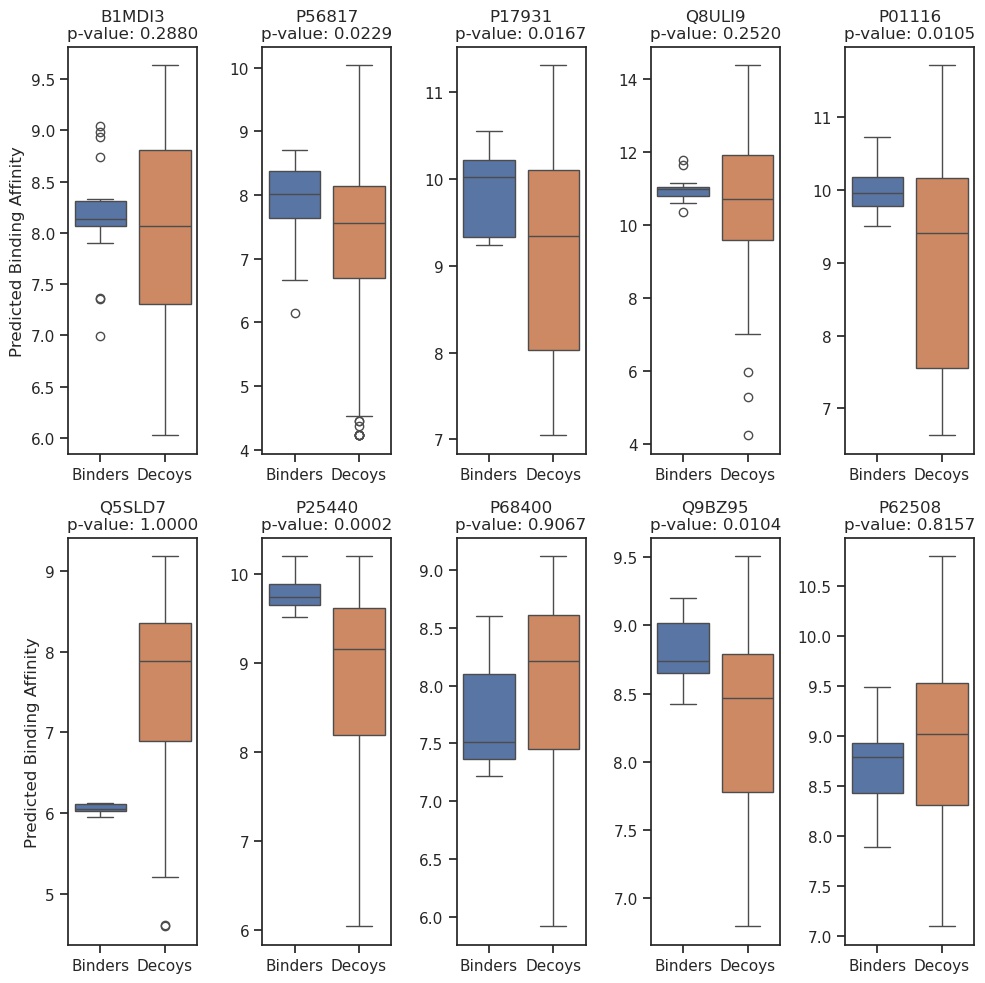}
    \caption{Predicted binding affinities of true binders and decoys for the five proteins in the PDBBind test set with the highest number of binders, using the protein structure of the third-lowest affinity binder.}
    \label{fig:ba3}
\end{figure}

\begin{figure}[htpb]
     \centering
     \begin{subfigure}[b]{0.32\linewidth}
         \centering
         \includegraphics[width=\linewidth]{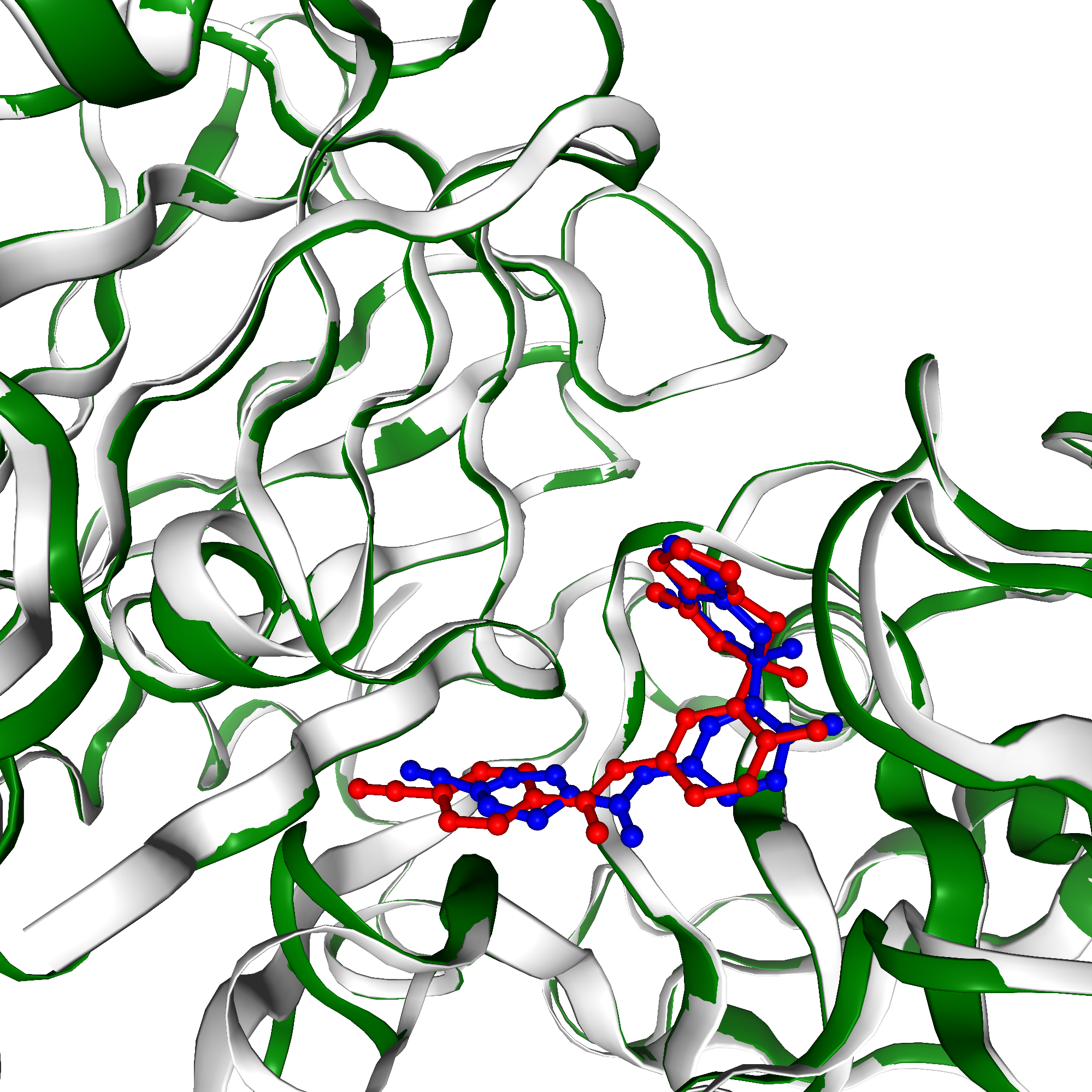}
         \caption{P56817, 0.8\AA.}
     \end{subfigure}
     \hfill
     \begin{subfigure}[b]{0.32\linewidth}
         \centering
         \includegraphics[width=\linewidth]{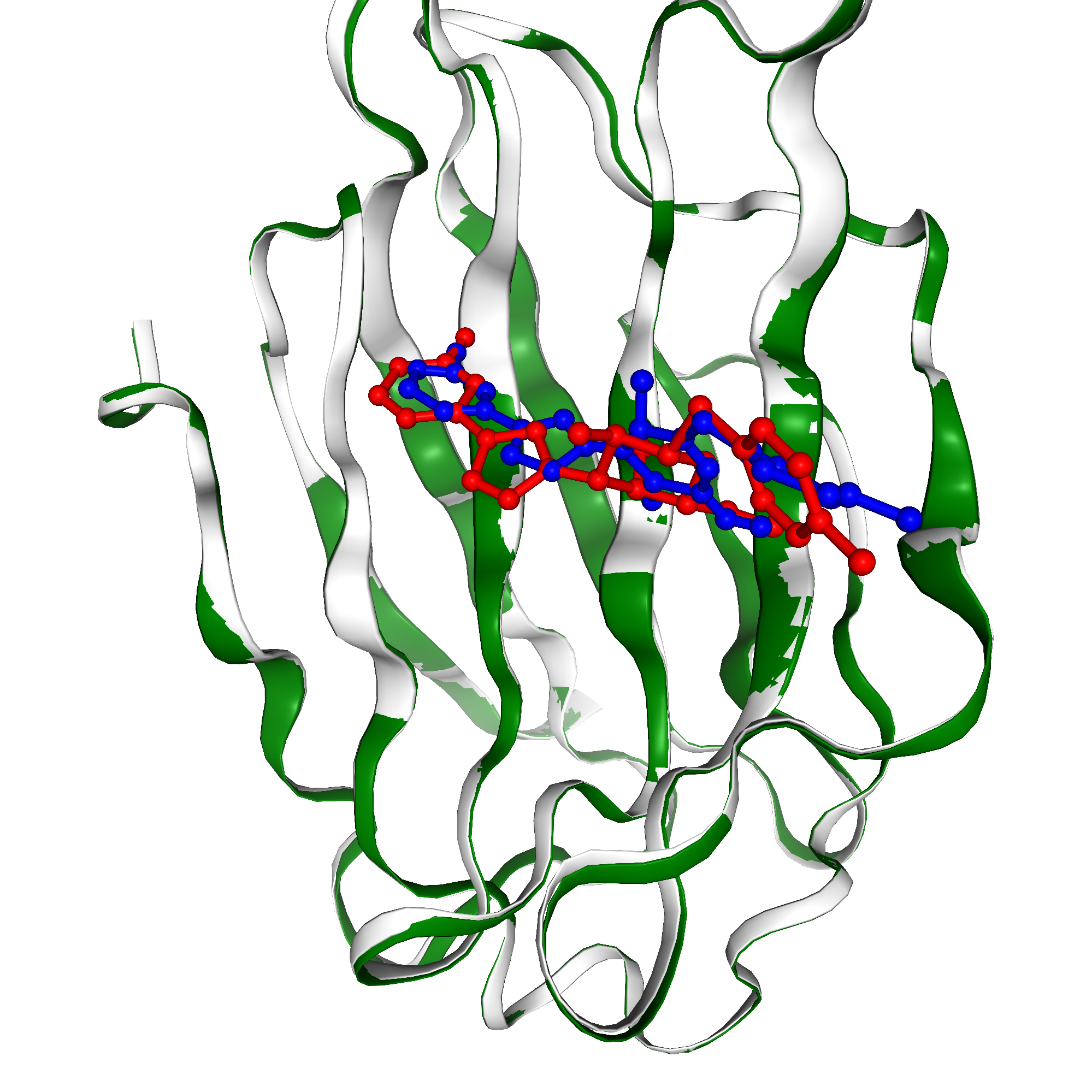}
         \caption{P17931, 1.1\AA.}
     \end{subfigure}
     \hfill
     \begin{subfigure}[b]{0.32\linewidth}
         \centering
         \includegraphics[width=\linewidth]{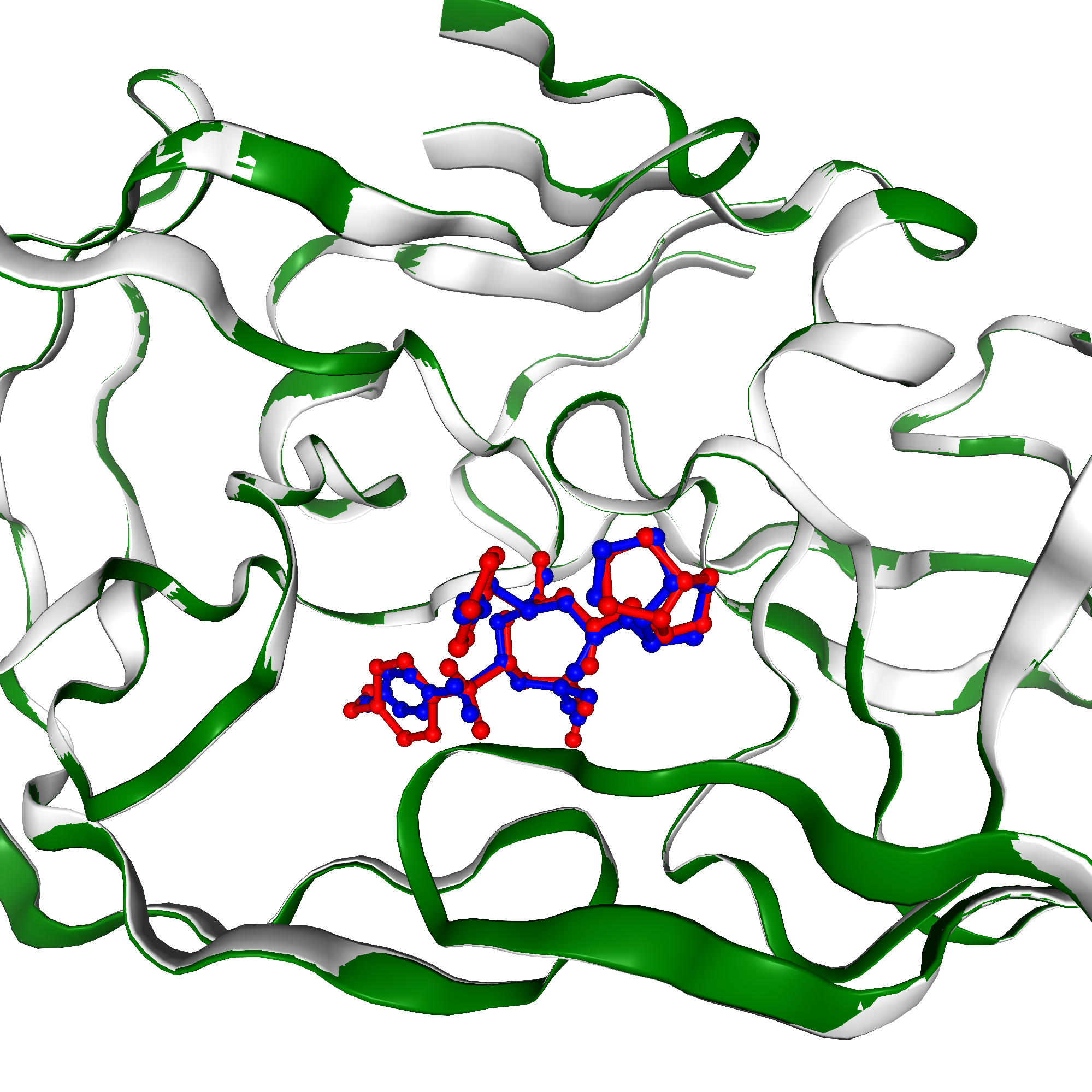}
         \caption{Q8ULI9, 0.7\AA.}
     \end{subfigure}
        \caption{Lowest RMSD \textsc{QuickBind} cross-docking predictions for P56817, P17931, and Q8ULI9. The ground-truth ligand and protein are shown in red and white, respectively. The input protein structure is shown in green and the \textsc{QuickBind} prediction is shown in blue.}
        \label{fig:crossdocking_preds}
\end{figure}

\clearpage

\section{Retrospective comparison with AF3}\label{sec:af3-comp}
In this section we discuss our observations on the differences introduced to \textsc{AF2} by \textsc{QuickBind} versus those made by \textsc{AF3}.

First, \textsc{AF3} no longer uses residue reference frames and eschews SE(3)-equivariance entirely. In early internal experiments on \textsc{QuickBind}, we similarly observed higher success rates when ligand reference frames were omitted. We opted to include them as the model would otherwise not be equivariant to the inputted global orientation of the protein-ligand complex, which is not desirable for docking applications. This decision was however driven by the fact that \textsc{QuickBind} uses an existing protein structure instead of predicting it from scratch. For a co-folding model, abandoning reference frames and SE(3)-equivariance is therefore consistent with our findings.

Second, \textsc{AF3} replaces the Evoformer with the Pairformer module. \textsc{QuickBind}'s modified Evoformer is architecturally a middle ground between the two. Similar to the Pairformer, it operates only on single and pair representations without column-wise attention. In the case of \textsc{QuickBind}, our design was informed by its use of an input protein structure, which obviated the need for a multiple sequence alignment and corresponding representation. In the Pairformer, the single representation does not update the pair representation via the outer product mean (OPM) module, and the update order of the single and pair representations is swapped. In \textsc{AF-Multimer} \cite{AFMultimer}, the OPM is moved to the beginning of the Evoformer. We found the original OPM position to be more optimal (see section~\ref{sec:hyperparamsscreen}), but did not try omitting it, or swapping the single and pair update order.

Third, \textsc{AF3} uses larger crop sizes than \textsc{AF2} and \textsc{AF-Multimer}. \textsc{AF3} is initially trained with a crop size of 384 then finetuned in two stages with crop sizes of 640 and 768, whereas \textsc{AF2} and \textsc{AF-Multimer} were trained with crop sizes of 256 and 384 then finetuned on crop sizes of 384 and 640, respectively. We initially trained \textsc{QuickBind} on 256 residue crops, then finetuned it using 512 residue crops. We found finetuning with larger crops to be important for model performance, observing consistent improvements as crop sizes increased. Given \textsc{AF3}'s training procedure, this suggests that \textsc{QuickBind} would benefit from additional finetuning stages with incrementally larger crops.

Fourth, \textsc{AF3} randomly chooses from contiguous, spatial, and spatial interface cropping. \textsc{QuickBind}'s cropping strategy, binding site cropping, can be considered a compromise between \textsc{AF3}'s two spatial cropping strategies and contiguous cropping. We tested spatial and contiguous cropping but found that binding site cropping leads to better results. Unlike \textsc{QuickBind}, \textsc{AF3} cropping is applied to all tokens such that the model may only see parts of the ligand, while in \textsc{QuickBind} only the protein is cropped.

Fifth, \textsc{AF3} contains a distogram head similar to the one in \textsc{AF2}, including a minimum distance bin of 2\AA\, which is overly large for small molecules (C-C bonds are 1.54\AA\ long). This suggests it primarily benefits overall complex prediction and rough positioning of ligand atoms, consistent with our observation that a distogram head does not improve \textsc{QuickBind}'s performance.

Finally, in agreement with the fact that we found better results when not distinguishing different bond types in the ligand adjacency matrix, \textsc{AF3}'s bond features are binary.

\end{document}